\documentstyle[11pt]{article}
\normalbaselines \oddsidemargin 0.5cm \evensidemargin 0.5cm
\begin{document}

\begin{center}
{\LARGE {\bf Weyl geometry approach to describe planetary systems }}\\
\vspace{2cm} ${P.~ Moyassari^
\dag}\footnote{e-mail:~p-moyassari@cc.sbu.ac.ir.} \hspace{.4cm}
and \hspace{.4cm}
\vspace{1cm}{S.~Jalalzadeh^\dag}\footnote{e-mail:~s-jalalzadeh@cc.sbu.ac.ir.}$
\vspace{0.5cm}\\  {\small {\dag Department of Physics, Shahid
Beheshti University, Evin, Tehran 19839,  Iran.}}
\end{center}
\vspace{0.5cm} \vskip 1cm
\begin{abstract}
In the present work we show that planetary mean distances can be
calculated through considering the Weyl geometry. We interpret the
Weyl gauge field as a vector field associated with the hypercharge
of the particles and apply the gauge concept of the Weyl geometry.
The  results obtained are shown to agree with the observed orbits
of all the planets and of the asteroid belt in the solar system,
with some empty states.
\end{abstract}
\section{Introduction}
In recent years, some quantum mechanical approaches have been
presented to calculate the planetary orbits in solar system
\cite{1,2,3}. Most of these theoretical approaches are based on
self-similarity concept which emphasizes the presence of
quantization at different scales. These methods implies the
existence of a re-scaled Plank constant or fine structure constant
as $\hbar^*\approx 10^{42}$js whose physical origin is still
undetermined. In this work we proceed to consider the Weyl
geometry instead of Remannian and show that Weyl geometry
introduces automatically quantization conditions for the planetary
orbits. Furthermore, we present a different interpretation of the
Weyl gauge field, in that we associate it with the spin-one
hyperphoton coupled to hypercharge, rather than the charge. Using
this interpretation, we can give a physical meaning to the
re-scaled fine structure constant (or Plank constant) mentioned in
the other approaches. All results will be obtained with just one
input parameter, namely a fundamental radius around 0.04 AU which
is predicted by some authors and several planets have been
recently discovered orbiting at this radius in extra solar
systems. In this paper we first briefly review the structure of
Weyl geometry and present a new interpretation of the Weyl gauge
field. Then we calculate the spherical solution of gravitational
field equation in Weyl-Dirac theory in section two. The planetary
mean distances are calculated through applying the gauge concept
of the Weyl geometry in the last section.

\section{Weyl Geometry}
After Einstein put forth his general theory of relativity, which
provided a geometrical description of gravitation, Weyl proposed a
more general theory that also included a geometrical description
of electromagnetism. In the case of general relativity one has a
Riemannian geometry with a metric tensor $g_{\mu\nu}$. If a vector
undergoes a parallel displacement in this geometry, its direction
may change, but not its length. In Weyl geometry, there is a given
vector $k^\mu$ which, together with $g_{\mu\nu}$, characterizes
the geometry. For any given vector $\xi^{\mu}$ undergoing parallel
displacement in this geometry, not only the direction but also the
length $\xi$ may change and this change depends on $k^\mu$
 according to the relation

\begin{equation}
d\xi=\xi k_\mu dx^\mu  \hspace{1cm}or \hspace{1cm} \xi=\xi_0
e^{\int k_\mu dx^\mu},
\end{equation}
where $\xi_{0}$ is the length of original vector before
displacement. The change in the length of $\xi$ in going from one
point to another depends on the path followed, i.e., length is not
integrable. This mathematical theory of displacement provides a
great flexibility in the choice of the standard of length. Thus,
one can introduce an arbitrary standard of length, or gauge, at
each point
\begin{equation}
ds'=e^\lambda ds
\end{equation}
here $\lambda$ is an arbitrary function of the coordinates. To
preserve equation (1) under this gauge transformation, $k_{\mu}$
transforms according to
\begin{equation}
k_\mu\longrightarrow k'_{\mu}=k_\mu+\partial_{\mu}\lambda.
\end{equation}
Straightforward generalization of Einstein-Hilbert action to Weyl
geometry leads to a higher order theory \cite{4}. Dirac
 introduced a new action called Weyl-Dirac action, by including a
new scalar field avoiding the presence of higher order terms in
action.
 The Weyl-Dirac action is given by \cite{5}
\begin{equation}
\begin{array}{ll}\vspace{0.5cm}
S[ \phi , k_{\mu} ] = \frac{1}{2}\int d^{4}x \sqrt{-g} \{
\frac{1}{2}F_{\mu\nu}F^{\mu\nu} +R\phi^{2}+\alpha
\phi_{\mu}\phi^{\mu}+\\
\hspace{2cm}(\alpha-6)\phi^{2}k_{\mu}k^{\mu}+ 2(\alpha-6) \phi
k^{\mu}\phi_{\mu}
+\frac{\lambda}{2}\phi^{4}+6(\phi^{2}k^{\mu})_{;\mu}\}
\end{array}
\end{equation}
where $F_{\mu\nu} = k_{\nu,\mu} - k_{\mu,\nu}$. This action is
invariant under the gauge transformations:
\begin{equation}
\begin{array}{lll}
g_{\mu\nu}\longrightarrow
e^{2\lambda}g_{\mu\nu}\\ \phi\longrightarrow e^{-\lambda} \phi\\
k_\mu\longrightarrow k_\mu+\partial_{\mu}\lambda
\end{array}.
\end{equation}
Varying the action (4) with respect to $g_{\mu\nu}$, $k_{\mu}$ and
$\phi$ yields
\begin{equation}
G_{\mu\nu}=\phi^{-2} [T_{\mu\nu} + \tau_{\mu\nu}+ t_{\mu\nu}]
\end{equation}
\begin{equation}
-(F^{\nu\mu})_{;\mu}+(\alpha-6)(\phi^{2}k^{\nu}+\phi\phi^{\nu})=0\
\end{equation}
\begin{equation}
\alpha\phi^{\mu}_{;\mu} - R \phi -(\alpha-6)[\phi k_{\mu}k^{\mu}-
\phi k^{\mu}_{;\mu}]- \lambda \phi^{3}= 0 \label{v2}
\end{equation}
here
\begin{equation}\label{s1}
T_{\mu\nu}[\phi]=(-2+\frac{\alpha}{2})
g_{\mu\nu}\phi_{\alpha}\phi^{\alpha}+(2-\alpha)
\phi_{\mu}\phi_{\nu}-2g_{\mu\nu}\phi(\phi^{\alpha})_{;\alpha}+
2\phi\phi_{\mu;\nu}+\frac{\lambda}{4}g_{\mu\nu}\phi^{4}
\end{equation}
\begin{equation}\tau_{\mu\nu}=\frac{1}{4}F_{\alpha\beta}F^{\alpha\beta}g_{\mu\nu}-
F_{\mu\alpha}F_{\nu\beta}g^{\alpha\beta}
\end{equation}
and
\begin{equation}
t_{\mu\nu}= (\alpha-6)[ \phi^{2}(- k_{\mu}k_{\nu}+
\frac{1}{2}g_{\mu\nu} k^{\alpha} k_{\alpha})-\phi(
k_{\mu}\phi_{\nu}+ k_{\nu} \phi_{\mu}-
k_{\alpha}\phi^{\alpha}g_{\mu \nu})]\label{s3}.
\end{equation}
Equation (3) is a familiar transformation that one has in the
Maxwell theory for the electromagnetic vector potential, and it
may lead to the identification of $k_\mu$ as the electromagnetic
vector field. But, it was shown that the usual spin $1/2$ fermions
(such as quarks or electrons) and the vector particles (such as
the Yang-Mills field or photons) cannot couple in minimal form to
the Weyl gauge field $k_\mu$ \cite{6}. Also, on the basis of the
Ehlers-Pirani-Schild axioms, $F_{\mu\nu}$ must be viewed as the
formal description of a phenomenon which has not been observed in
nature \cite{7}. In this paper we present a different
interpretation of the Weyl gauge field . We assume that the vector
field $k^\mu$ is associated with the spine-one hyperphoton coupled
to hypercharge, rather than the charge (hypercharge is defined as
the sum of the baryon number and the strangeness.). Since the
discovery that $K_{2}^0$ as the long lived component of the
$K^{0}$ decays into two $\pi$ mesons (a CP violating interaction),
it was pointed out that the observed effect can be interpreted by
supposing a new long-range interaction between $K$ meson and our
galaxy \cite{8}. It was postulated that this interaction can be
mediated by a vector field analogous to the electromagnetic field
but coupled to hypercharge rather than the charge. This
interaction produces a potential energy equal in magnitude and
opposite in sign for particles and antiparticles \cite{8}. The
configuration for this vector field analogous with the
electrodynamic is
\begin{equation}
A_0=gH/r , \hspace{.6 cm} A_i=0
\end{equation}
Here $g$ is the hypercharge coupling constant and $H$ is the
number of nuclei in the central mass. Therefore, we assume that
the Weyl gauge field $k_\mu$ is the vector field analogous whit
this large range potential.

\section{Spherical solution}

In this section we attempt to obtain  spherical solution of the
field equation (6). We assume the factor $(\alpha-6)$ to be small
\cite{9}. So, we neglect the terms including this factor. Let us
take the line element as
\begin{equation}
ds^2 = -e^{\nu}dt^2 + e^{\lambda}dr^2 + r^2d\Omega^2.
\end{equation}
where $\nu$ and $\lambda$ are functions of $t$ and $r$. The
absence of a generalized Birkhoff theorem in Weyl space allows the
existence of the dynamical solutions in the spherical symmetric
model under consideration \cite{10}. Solving the equation (7)
leads to
\begin{equation}
k_{0,r}=\frac{\gamma(t)e^{(\nu+\lambda)}}{r^2}
\end{equation}
where $\gamma(t)$ is an arbitrary function of time arising from
integration. The field equation (6) together with  (14) gives
\begin{equation}
e^{-\lambda}\left(-\frac{\lambda_{,r}}{r} + \frac{1}{r^2}\right) -
\frac{1}{r^2} = \frac{1}{\phi}e^{-\nu} ( \phi_{,t}\lambda_{,t} +
\frac{3\phi^2_{,t}}{\phi^2})-\frac{\gamma^2}{r^4\phi^2}
\end{equation}
\begin{equation}
e^{-\lambda}(\frac{\nu_{,r}}{r} + \frac{1}{r^2} ) - \frac{1}{r^2}
= \frac{1}{\phi}e^{-\nu} ( \phi_{,tt} - \frac{\phi^2_{,t}}{\phi^2}
- \phi_{,t}\nu_{,t} )-\frac{\gamma^2}{r^4\phi^2}
\end{equation}
\begin{equation}
-\frac{\lambda_{,t}}{r} = \frac{\phi_{,t}\nu_{,r}}{\phi}.
\end{equation}
Here the subscripts denote partial derivatives. In view of the
forms of the left sides of equations (15,16), we assume that
\begin{equation}
e^\nu=f(t)e^{-\lambda}
\end{equation}
Making use of (17) and (18), one finds that equations (15) and
(16) become identical provided $f(t)$ satisfies
\begin{equation}
f_{,t}/f=2\phi_{,tt}/\phi_{,t} -4\phi_{,t}/\phi
\end{equation}
this gives
\begin{equation}
f=\frac{(\phi_{,t})^2}{\alpha^2\phi^4} \hspace{.6 cm}
(\alpha=const).
\end{equation}
After some manipulation we obtain the generalization of the result
derived by Rosen \cite{9}, as
\begin{equation}
e^\nu=(\phi^2_{,t}/\alpha^2\phi^4)e^{-\lambda}
\end{equation}
with
\begin{equation}
\begin{array}{ll}\vspace{0.5cm}
e^{- \lambda} = \frac{1}{2} - \frac{m_{0}}{\phi
r}+\frac{b^2}{\phi^2r^2} +\Delta
\\ \Delta = [ ( \frac{1}{2} -\frac{m_{0}}{\phi r}+\frac{b^2}{\phi^2r^2})^2 + \alpha^2
\phi^2r^2]^{\frac{1}{2}}
\end{array}
\end{equation}
Here we assume that $\gamma(t)=b\phi^{-1}$ in which $b$ is a
constant such as $m_{0}$. One can interpret $m_0$ and $b$ as the
mass and hypercharge of the central mass respectively. The above
solution is valid for any gauge function $\phi(t)$. We choose the
Einstein gauge $( ds^2 \rightarrow d\bar{s}^2 = \phi^2ds^2 , \phi
\rightarrow \bar{\phi} = 1 )$, thus one gets
\begin{equation}
d\bar{s}^2 = -e^{-\lambda}dT^2 + e^{2\alpha T}( e^{\lambda}dr^2 +
r^2d\Omega^2 )
\end{equation}
where
\begin{equation}
dT = \frac{\phi_{,t}}{\alpha \phi}dt = \frac{d\phi}{\alpha \phi}
\end{equation}
so that, with a suitable choice of the origin of $T$ in the
natural units ($c=\hbar=1$), we have
\begin{equation}
\phi=e^{\alpha T}
\end{equation}
From the coefficient of $T_{\mu\nu}$ in equation (6) one can
postulate that the gravitational coupling $G$ behaves as
\begin{equation}
G=G_0\phi^{-2}.
\end{equation}
where $G_0$ is the gravitational constant in the present epoch.
Comparing this equation with equation (25), shows
\begin{equation}
G\approx G_0(1-2T/T_0)
\end{equation}
where $T_0$ is the present age of the universe and $T$ is the time
elapsed from the present epoch. One can see that the spherical
solutions of the Weyl-Dirac theory leads to the time varying
gravitational constant in accordance with \cite{11}.
\section{Orbit of planets}
We proceed now to show that planetary mean distances can be
calculated in a Weyl spacetime. We investigate the motion of a
planet in the gravitational field of the sun and apply to it the
gauge concept of the Weyl geometry. This method was also used for
the motion of an electron in the electromagnetic field of a proton
in a hydrogen atom \cite{12}. These considerations provides for
obtaining the Bohr radii of quantum theory. Thus, the Weyl
geometry introduces automatically quantization conditions for the
orbits in the hydrogen atom. We now apply this model to the solar
system. In this scale one can make the reasonable assumption that
there is no pure electric charge, but instead, the hypercharge
will be dominant. We investigate the effects of the hypercharge of
the sun on the motion of the planets through considering the gauge
field $k_\mu$ as the hypercharge vector field induced by the long
range potential. So, the non-vanishing component of $k_\mu$ is
$k_0=gA_0=g^2H/r$. Since we interpret $b$ as the hypercharge
 of the central mass in the equation (22) one can compute
this constant in terms of the hypercharge of the sun as $(b = g^2
H)$ which according to Bell's estimation is approximately
$b\approx 10^9$ \cite{8}. Now, we first calculate the
gravitational potential of the sun by using the spherical solution
of the Weyl-Dirac gravitational equation. Let $g_{00}$ be the
component of the metric (13) in the conventional units in the
present epoch, then we get from (22)
\begin{equation}
g_{00}=\frac{1}{2}-\frac{G_0M}{r}+(\frac{g^2H}{r})^2+\Delta
\end{equation}
where
\begin{equation}
\Delta=[(\frac{1}{2}-\frac{G_0M}{r}+(\frac{g^2H}{r})^2)^2+(\frac{r}{T_0})^2]^{1/2}
.
\end{equation}
and $M$ is mass of the sun. It is obvious that the effect of the
term proportional to $1/r^2$ in the large scale is ignorable
compared to the term proportional to $1/r$. Thus, by expanding the
above equation to the first order we obtain
\begin{equation}
g_{00}=1-\frac{2G_0M}{r}+ \frac{2G_0M
r}{T_0^2}+\frac{r^2}{T_0^2}-8(\frac{G_0M}{T_0})^2.
\end{equation}
One can see that the existence of the linear terms in the
potential are due to the variable gravitational coupling, in
accordance with \cite{11}. Now, we consider the motion of a planet
in the gravitational potential of the sun, then for a given orbit
we have
\begin{equation}
\frac{G_0M}{r^2}+\frac{G_0M}{T_0^2}+\frac{r}{T_0^2}=\frac{v^2}{r}
\end{equation}
We assume that the orbits of the planets are integrable in the
Weyl geometry. It means that the frequency of spectral lines
evidently does not depend on the history of the radiating matter.
\begin{equation}
\exp(\int_0^\tau k_0 d\tau')=1
\end{equation}
here $\tau$ is the period of motion of a planet around it's orbit.
This relation seems like a quantization condition. By using
equations (31) and (32) the possible radius $r$ for the orbits
around which the length scale is preserved in the solar scale are
given by
\begin{equation}
r=\frac{n^2G_0M}{g^4H^2}
\end{equation}
which corresponds to the derived expression for the radii of
orbits by Agnese and Festa \cite{1}, but there is a difference.
There is a constant parameter in the result derived in \cite{1}
which is called re-scaled Planck constant \cite{3} or re-scaled
fine structure constant \cite{1} and there is no physical origin
accounting for them. Here the term $g^2H$ plays the role of the
re-scaled fine structure constant considered in \cite{1}. As one
can see in the present model the Weyl geometry can provide a
physical meaning for it.

Now, we assume a fundamental radius $r_1=0.04AU$ for $n=1$, as was
predicted by Nottale \cite{2} and also by Agnese \cite{13}.
Several extra solar planets recently discovered lie at this
distance from their star \cite{14}. Using this assumption, we can
estimate the hypercharge coupling constant $g$, as $g^2=10^{-53}$
which is roughly related to the Bell's estimation \cite{8}. By
this consideration we can obtain a sequence of values that assures
very well the observed values of the orbital radii in our solar
system. For $n$ running from 3 to 6, we obtain orbital radii of
Mercury, Venus, Earth and Mars; for $n=8$ we have the asteroid
Ceres, and for $n=11, 15, 21, 26$ and 30, the orbital radii of
Jupiter, Saturn, Uranus, Neptune and Pluto follow, in agreement
with the observed values. The problem here, as we can see, is that
we fall into a lot of empty orbital positions which are predicted
by the formula $r=n^2r_1$, but not occupied by any observed body,
particularly for large values of $n$.

\section{Conclusion}

In this paper we have presented a new model for obtaining the
orbits of the planets and the asteroid belts. Our approach is in
the direction of many recent works that have been done in
geometrization of quantum processes. In this manner we have shown,
using Weyl-Dirac approach to  gravity that one can describe solar
scale quantization. A major ingredient of this model is the Weyl
gauge field which can be interpreted as a gauge field describing
the hypercharge of particles. This is the dominant field at large
scales and would provide a physical meaning to the re-scaled
Planck constant in solar and also in cosmic scales.
\noindent\vspace{7mm}\\
{\bf Acknowledgments}\noindent \\
 We would like to thank H. Salehi
for introducing us to this problem and H. R. Sepangi for his
careful reading.

\end{document}